\documentclass[
    ,final            
  ,numberedheadings 
  ]
  {aipproc}

\layoutstyle{8x11single}
\usepackage{epsfig}

\newcommand{\dtres}{d^{\hspace{0.1mm} 3}\hspace{-0.5mm}}

\newcommand{\sqd}{\sqrt{2}}

\newcommand{\be}{\begin{eqnarray}}
\newcommand{\ee}{\end{eqnarray}}
\newcommand{\nn}{\nonumber}
\newcommand{\ddn}{D^0\bar D^{*0}}
\newcommand{\ddc}{D^+ D^{*-}}
\newcommand{\ddi}{D\bar D^{*}}
\newcommand{\kvec}{\vec{k}}
\newcommand{\pvec}{\vec{p}}

\newcommand{\ppvec}{\vec{p}^{\;\prime}}

\newcommand{\rvec}{\vec{r}}

\newcommand{\Eq}[1]{Eq.~(\ref{#1})}

\newcommand{\psihat}{\hat \psi}

\begin{document}

\title{Isospin breaking effects in the
dynamical generation of the $X(3872)$}


\author{D. Gamermann}{
  address={Instituto de F\'isica corpuscular (IFIC), Centro Mixto
Universidad de Valencia-CSIC,\\ Institutos de Investigaci\'on de
Paterna, Aptdo. 22085, 46071, Valencia, Spain}
}

\author{J. Nieves}{
  address={Instituto de F\'isica corpuscular (IFIC), Centro Mixto
Universidad de Valencia-CSIC,\\ Institutos de Investigaci\'on de
Paterna, Aptdo. 22085, 46071, Valencia, Spain}
}

\author{E. Oset}{
  address={Departamento de F\'isica Te\'orica and IFIC, Centro Mixto
Universidad de Valencia-CSIC,\\ Institutos de Investigaci\'on de
Paterna, Aptdo. 22085, 46071, Valencia, Spain}
}

\author{E. Ruiz Arriola}{
  address={Departamento de F\'isica At\'omica, Molecular y Nuclear,
Universidad de Granada, E-18071 Granada, Spain}
}

\begin{abstract}
 We have studied isospin breaking effects in the X(3872) resonance and found a natural explanation for the branching fraction of the X decaying to $J/\psi$ with two and three pions being close to unit. Within our framework the X(3872) is a dynamically generated resonance in coupled channels. We also study the relationship between the couplings of the resonance to the coupled channels with its wave function, which further helps us to understand the isospin structure of the resonance. 
\end{abstract}

\maketitle


\section{Introduction}

The $X(3872)$ resonance has been first observed by the Belle collaboration \cite{belle} and later confirmed
by other four experiments the CDFII, D0 and BaBar collaborations \cite{cdf,d0,babar}. It has first been seen in
the decay channel $J/\psi \pi^+\pi^-$ and the analysis of the $\pi\pi$ spectrum indicates that this meson pair
comes from a $\rho$ meson. There has been no observation of a charge partner for the $X(3872)$ state, which indicates
an isoscalar nature for it. Later on the $X(3872)$ has been observed in another two decay channels \cite{bellegj}: 
The channel $J/\psi \gamma$ that fixes the C-parity of the state as positive and the channel $J/\psi\pi^+\pi^-\pi^0$
where the three pion spectrum shows a peak for an $\omega$ resonance. The measured branching fraction

\be
\frac{{\cal B}(X\rightarrow J/\psi\pi^+\pi^-\pi^0)}{{\cal B}(X\rightarrow J/\psi\pi^+\pi^-)}&=&
1.0\pm0.4\pm0.3\textrm{ ,} \label{branch}
\ee 
raises a problem, since it seems to indicate a huge isospin violation in the decays of the $X(3872)$, since the case of
$\rho$ production (two pions) has isospin 1 while the $\omega$ has isospin 0. A careful analysis of the data on this state
has been done in \cite{xqn}, concluding that its quantum numbers must correspond to $J^P=1^{++}$ or $J^P=2^{-+}$.

From the theoretical point of view the $X(3872)$ mass is very close to the $\ddn$ threshold and it has been suggested
\cite{tornq,swanson,dong,meuax} that this state is a $\ddi$ molecule. There has been though some discussion on
whether the $\ddc$ component is important in the structure of the state, since this component is bound by about 8 MeV,
while the neutral component is bound by less than 1 MeV \cite{braaten,meutwox}.

In this work we explain how the branching fraction of \Eq{branch} is naturally explained for a dominant isospin 0
nature of the $X(3872)$. We also stress the importance of taking into account the charged $\ddi$ components in the
structure of the $X$ even though the neutral wave function extends much further in space than the charged one. This
is a consequence of the fact that only the values of the wave function around the origin are important in the
description of short range processes as the decays to $J/\psi$ with two or three pions.


\section{The Decays of the $X(3872)$ into two and three Pions}

We follow here the approach of \cite{meutwox} where the $X(3872)$ is generated dynamically within coupled channels.
As noted in \cite{meuax,meutwox} only the $\ddc$ and $\ddn$ components of the state provide sizeable couplings
and therefore we consider only these two channels in the approach.

The model is based on the calculation of the unitarized transition matrix (T-matrix). For the tree level interaction
of the mesons, Lagrangians based on the $SU(4)$ extensions of the hidden gauge Lagrangians are used. The $SU(4)$
symmetry is broken in order to take into account the bigger mass of the charmed and hidden charmed mesons.
The tree level amplitudes are used as kernel to solve a scattering equation that in an on-shell approach
takes the form 

\be
T&=&(1-VG)^{-1}V \label{bseq}
\ee
where $T$ is the T-matrix, $V$ is a matrix containing the tree level amplitudes projected in s-wave and $G$ is a
diagonal matrix with each diagonal element containing the two particle propagators for each channel.

The loop functions in $G$ are divergent and can be regularized using dimensional regularization, which gives a
subtraction constant $\alpha$ as a free parameter that can be used to adjust the pole generated in the T-matrix of
\Eq{bseq} to its right position, fitting in this way the $X(3872)$.

Close to the pole the T-matrix can be written in the separable form

\be
T_{ij}&=&\frac{g_i g_j}{s-s_{pole}}\label{coup}
\ee
where $s$ is the invariant mass squared, $s_{pole}$ is the pole position and $g_k$ are the couplings to each channel.
Using \Eq{coup} one can calculate the couplings of the resonance to each channel by calculating the residues of the
T-matrix at the pole position. By using physical masses for the charged and neutral components of the $D$-meson doublet
members, one finds a difference of at most 1.5\% in the couplings of the $X$ to the charged and neutral channels. This
is a tiny difference which we will neglect in what follows.

In order to calculate the ratio of the decays into $J/\psi\rho$ and $J/\psi\omega$ one has to take into account the
coupling first of the $X$ to a $\ddn$ or $\ddc$ and then the coupling of the $\ddi$ components to $J/\psi\rho$ or
$J/\psi\omega$. In the hidden gauge formalism this former coupling is equal for both cases. What differentiates the
situation with the $\rho$ meson from the situation with the $\omega$ mesons is that in the case of the $\rho$ which has isospin 1
the loops with neutral and charged $\ddi$ mesons interfere destructively while in the case of the $\omega$ with isospin 0
the interference is constructive. As a result the ratio of the coupling squared of the $X$ to $\rho$ and $\omega$
is given by

\be
R_{\rho/\omega}&=&\frac{G_{11}-G_{22}}{G_{11}+G_{22}}=0.032
\ee

The value 0.032 was obtained in the limit of 0 binding energy with the loops calculated with dimensional regularization.

In order to calculate the full ratio of the branching fractions one still has to take into account the phase space available
for the $\rho$ to decay into two pions and the $\omega$ into three pions. For that we integrate the spectral function 
of each meson taking into account the much bigger width of the $\rho$ and the kinematical constrains. The result
obtained is

\be
\frac{{\cal B}(X\rightarrow J/\psi\pi^+\pi^-\pi^0 )}{{\cal B}(X\rightarrow J/\psi\pi^+\pi^- )}&=&1.4 \label{res14}
\ee
which is compatible with the value $1.0\pm0.4$ from experiment \cite{bellegj}. One sees that although the couplings
of the $X$ to the isospin 1 state $J/\psi\rho$ is suppressed in relation to the isospin 0 state $J/\psi\omega$ by a 
factor 30, the much bigger phase space for the $\rho$ to decay than for the $\omega$, compensates this suppression and
brings the branching fraction to a value close to the observed experimentally.


\section{the Wave Functions of the $X(3872)$}

To calculate the wave functions of the $X$ resonance we start from a phenomenological potential that in momentum space
reads:

\be
\langle \ppvec|V|\vec{p}\,\rangle \equiv& V(\ppvec,\vec{p}\,)=&
v \,\Theta(\Lambda-p)\Theta(\Lambda-p^\prime) \nn\\
v&=&\left(\begin{tabular}{cc} ${\hat v}$ & ${\hat v}$
  \\ ${\hat v}$ &${\hat v}$ \end{tabular}\right)\label{potte}
\ee
where the cut off $\Lambda$ is fixed in order to find the resonance with the appropriate mass. To do that we write the
T-matrix and the Lippmann-Schwinger equation:

\be
\langle \vec{p}\,|T|\ppvec\rangle&\equiv&
T(\ppvec,\vec{p}\,)= t \, \Theta(\Lambda-p)\Theta(\Lambda-p^\prime) \\
t&=&(1-vG)^{-1}v
\ee
the coupled channels loop function $G$ here is a diagonal matrix given by:

\be 
\hspace{-0.3cm}G &=&\left(\begin{tabular}{cc} $G_{11}$ & 0
  \\ 0 & $G_{22}$ \end{tabular}\right), \quad
G_{ii}=\int_{p<\Lambda}
\frac{\dtres p}{E-M_i-\frac{\vec{p}^{\,2}}{2\mu_i}}\label{eq39}
\ee

The poles of the T-matrix are given by the equation det$(1-vG)=0$. The value of Lambda is chosen so that this equation
is satisfied for the loop calculated at the energy where we want to have the $X$ state bound.

We write then the Schor\"odinger equation for the potential:

\be
(H_0+V)|\psi\rangle&=&E_\alpha|\psi\rangle
\ee
where $H_0$ is the free Hamiltonian and $E_\alpha$ is the energy of the resonance. Projecting this equation in momentum
space one obtains a set of two coupled equation for the wave function in each channel:

\be
\langle \pvec\,|\psi_1\rangle &=&{\hat v}\frac{\Theta(\Lambda-p)}{E_\alpha-M_1-\frac{\vec{p}^{\,2}}{2\mu_1}}
\int_{k<\Lambda}\dtres k\, \left(\langle \kvec|\psi_1\rangle
+\langle \kvec|\psi_2\rangle \right) 
\label{eq44a} \\
\langle \pvec\,|\psi_2\rangle &=&{\hat v}\frac{\Theta(\Lambda-p)}{E_\alpha-M_2-\frac{\vec{p}^{\,2}}{2\mu_2}}
\int_{k<\Lambda}\dtres k \, \left (\langle \kvec|\psi_1\rangle +\langle \kvec|\psi_2\rangle \right) \label{eq44b}
\ee
The two integrals in the right hand side of the equations are constants that can be calculated by normalizing the wave
function.

The couplings of the state to the channels can again be calculated from the residues of the T-matrix:

\be
g_1^2=g_2^2 \equiv  g^2&=&\lim_{E\rightarrow E_\alpha}(E-E_\alpha)t_{ij}
=-\left.
\left(\frac{dG_{11}}{dE}+\frac{dG_{22}}{dE}\right)^{-1}\right|_{E=E_\alpha}\label{eqres}
\ee
In the limit when $E_{B1}^\alpha\rightarrow 0$, we have
$\left. \frac{dG_{11}}{dE}\right |_{E=E_\alpha}\rightarrow \infty$ and we find
\be
g_1^2=g_2^2&\sim &\frac{\gamma_1}{4\pi^2\mu_1^2}, \qquad
E_{B1}^\alpha\rightarrow 0   \label{eq40}
\ee
where
\be 
\gamma_i &=&\sqrt{2\mu_i E_{Bi}^\alpha} \\ 
E_{Bi}^\alpha &=& M_i-E_\alpha 
\label{eq:gamma-th}.
\ee 
and the $M_i$ is the value of the threshold for channel $i$.

As shown in \cite{meuwave} working out \Eq{eq44a}, \Eq{eq44b} and \Eq{eqres} one obtains 

\be
g\,G_{11}^\alpha=\int_{p<\Lambda} \dtres p\langle \pvec\,|\psi_1\rangle  \label{eq51a}\\
g\,G_{22}^\alpha=\int_{p<\Lambda} \dtres p\langle \pvec\,|\psi_2\rangle  \label{eq51b}
\ee
and we can recognize that the integrals in these equations are the Fourier transform of the momentum wave function
for $\vec{r}=0$. So we can rewrite \Eq{eq51a} and \Eq{eq51b} as

\be
gG^\alpha_{11}=&(2\pi)^{3/2}\psi_1(\vec{0}\,)=&\hat{\psi_1} \label{eq53a} \\
gG^\alpha_{22}=&(2\pi)^{3/2}\psi_2(\vec{0}\,)=&\hat{\psi_2} \label{eq53b} 
\ee

These equations show that the couplings of the the resonance to its channels are proportional to the value of the wave
function at the origin (denoted as $\psihat$).

Had we chosen another kind of form factor for regularize the potential in \Eq{potte} we would arrive at the same
equations, but with a slightly different definition of $\psihat$:

\be
{\hat \psi}_i&=&\int \dtres k f(\kvec)\langle\kvec|\psi_i\rangle \label{psihatn}
\ee
where $f(\kvec)$ is a form factor substituting the $\Theta(\Lambda-k)$ in \Eq{potte}. This new form factor has also a
scale $\Lambda$ that in coordinate space represents a range about $1/\Lambda$. Since a typical cut off of 700 MeV
represents in coordinate space a range of about 0.3 fm, $\psihat$ represents here a smeared value of the wave
function around the origin.

In Table \ref{tabcomp} we show results for potentials with three different form factors. The wave functions for each
kind of potential are slightly different but the values of $\psihat_i$ and the ratio of these values do not change much,
since these are the quantities that carry the information about the short range physics.

\begin{table}
\caption{Comparative results for different potentials for a $\ddn$
  binding energy of 0.1 MeV.} \label{tabcomp}
\vspace{0.5cm}
\begin{tabular}{c|c|c|c|c|c|c|c}
Form & $\Lambda$ & $g^{\textrm{FT}}$ & $\psi_1(\vec{0}\,)/\psi_2(\vec{0}\,)$ & $\psihat_1/\psihat_2$ & $\psihat_1$& $\psihat_2$& $R_{\rho/\omega}$ \\
Factor & [MeV] & [MeV] & & & & &\\
\hline
\hline 
Sharp & 653 & 3202 & 1.31 & 1.31 &3.29&2.50& 0.018 \\
\hline
Gauss & 731 & 3238 & 1.20 & 1.29 &3.30&2.56& 0.016 \\
\hline
Lorentz & 834 & 3254 & 1.17 & 1.28 &3.31&2.58& 0.015
\end{tabular}
\end{table}

\begin{figure}
\begin{tabular}{cc}
\includegraphics[width=6.0cm,angle=-0]{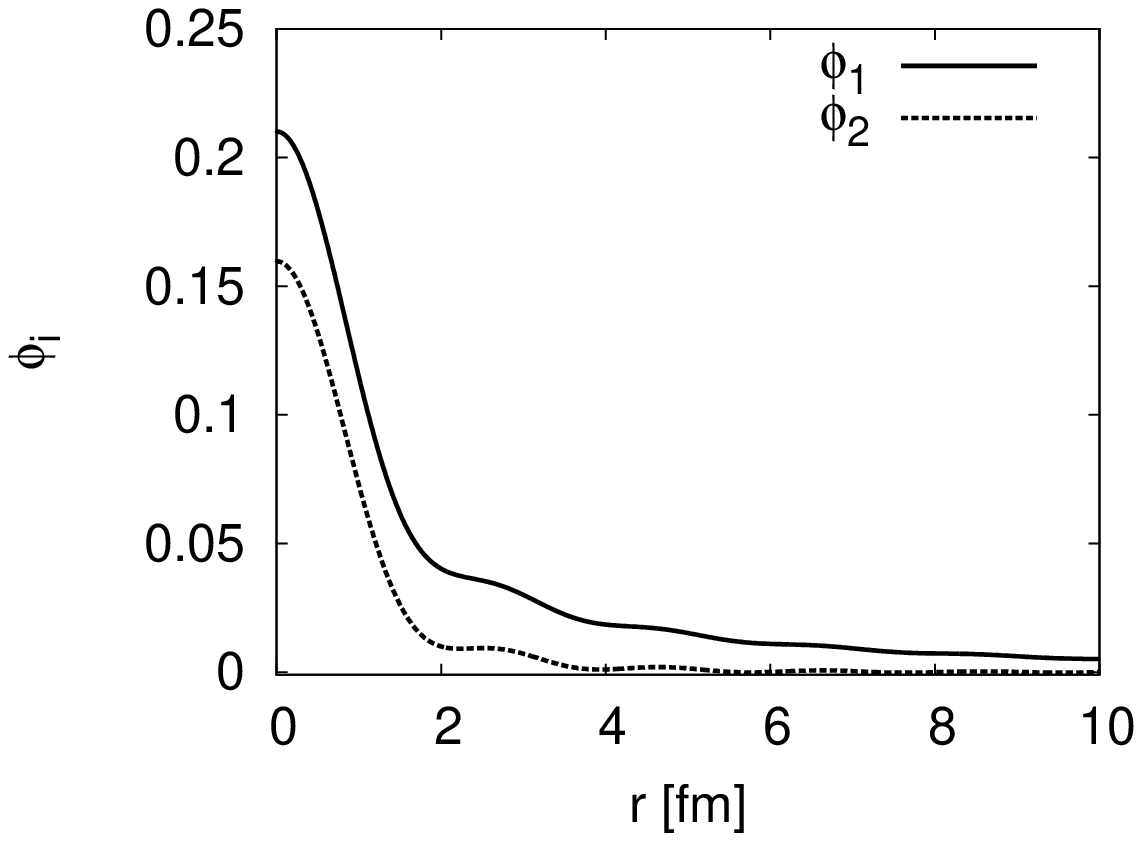} & \includegraphics[width=6.0cm,angle=-0]{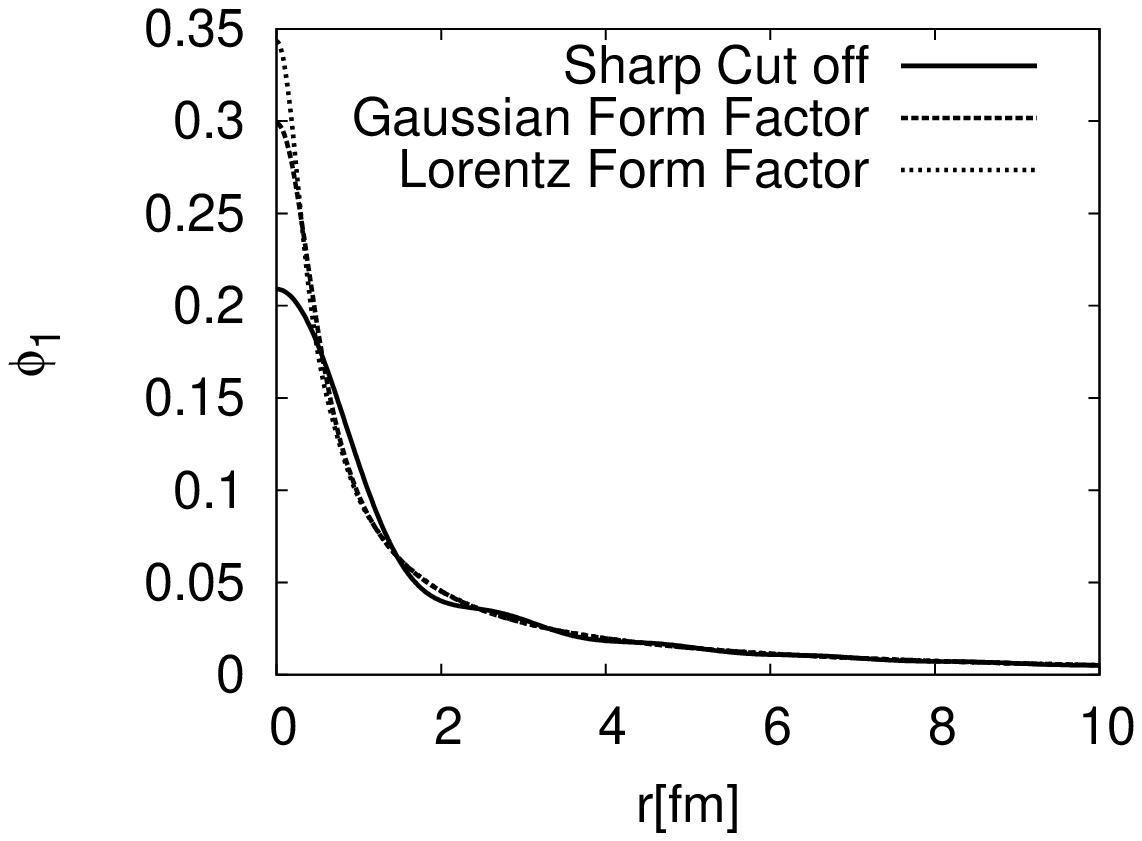}\\
\includegraphics[width=6.0cm,angle=-0]{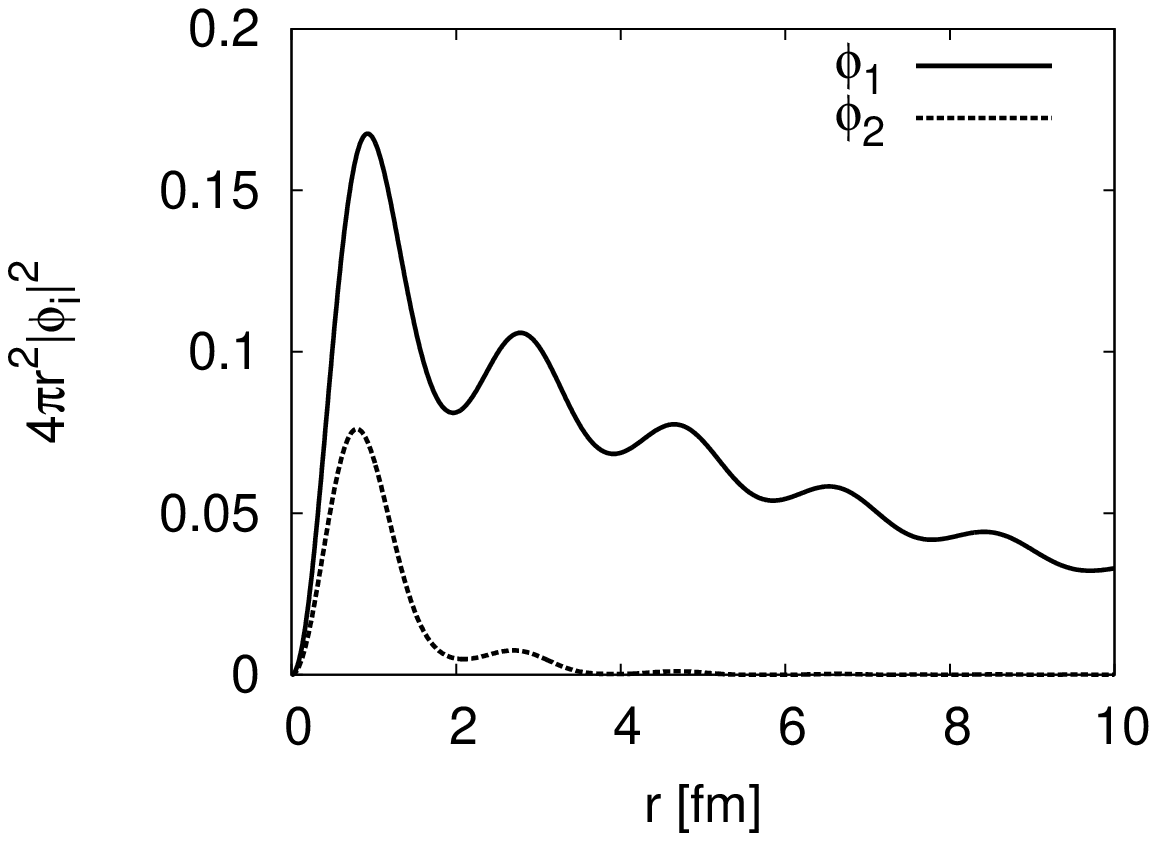} & \includegraphics[width=6.0cm,angle=-0]{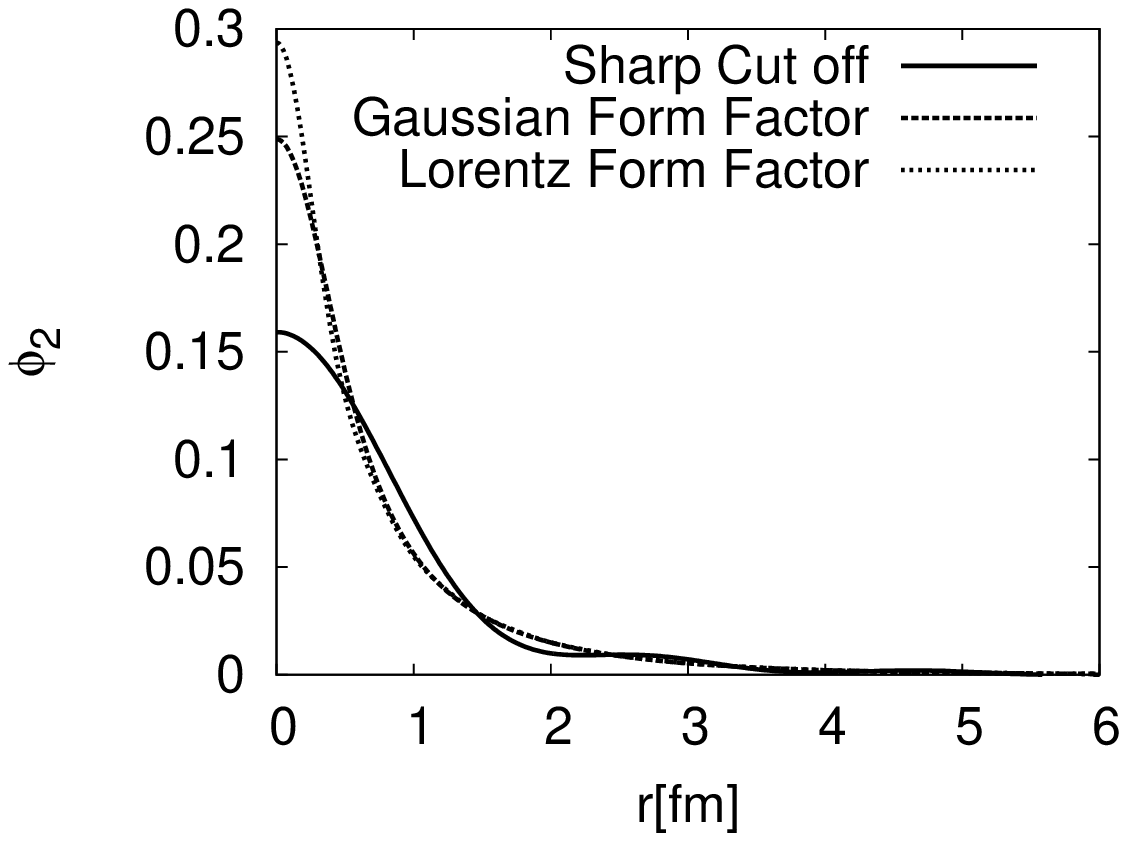} 
\end{tabular}
\caption{Neutral and charged Wave function components for a $\ddn$
  binding energy of 0.1 MeV. In the upper left panel one can see the value
  of the wave functions at the origin for both channels, in the lower left
  panel we plot the probability density for each channel. In the right panel we show plots of the wave functions
  for each channel calculated with the three different form factors in the potential.}
  \label{fig1}
\end{figure}

In Figure \ref{fig1} we show plots of the wave functions and probability densities for the sharp cut off potential
and plots comparing the wave functions for the three different kinds of form factors.

With regard to the isospin structure of the $X(3872)$ we can first write its wave function in terms of the charge basis
and then transform it to isospin basis:

\be
|X\rangle&=&\psi_1(\rvec)|\ddn\rangle+\psi_2(\rvec)|\ddc\rangle \\
|X\rangle&=&\frac{\psi_1(\rvec)+\psi_2(\rvec)}{\sqd}|\ddi_{I=0}\rangle+
\frac{\psi_1(\rvec)-\psi_2(\rvec)}{\sqd}|\ddi_{I=1}\rangle \label{xiso}
\ee
We see from \Eq{xiso} that the mixing between the two possible isospin states depends on the relative distance of the
two D mesons, $\rvec$. But in physical processes this mixing will be given by the values of the wave
function around the origin. Using the $\psihat_i$ values given in Table \ref{tabcomp} one sees that for short range
processes the contribution of the isospin 1 state is about 2\% and the $X$ can be regarded as an almost pure isospin 0
state.


\section{Conclusions}

We have studied the isospin structure of the $X(3872)$. In our model this state is generated dynamically in coupled
channels and is interpreted as an s-wave bound state of $\ddi$. The apparent isospin violation in the branching fraction
ratio of $X\rightarrow J/\psi\pi^+\pi^-$ and $X\rightarrow J/\psi\pi^+\pi^-\pi^0$ is naturally explained due to the
much larger phase space for the $\rho$ to decay into $\pi^+\pi^-$ than the $\omega$ to decay into $\pi^+\pi^-\pi^0$.

The mass of the $X(3872)$ is much closer to the $\ddn$ threshold than to the $\ddc$ which is bound by around 8 MeV.
As a consequence the wave function of the $X$ in the $\ddn$ channel extends much further on space than in the $\ddc$
channel. Nevertheless physical short range processes are dominated by the couplings of the state to each channel
and these couplings have a negligible isospin violation. In terms of wave functions this can be understood since
the couplings reflect an averaged value of the wave function around the origin and not in the whole space.


\begin{theacknowledgments}
 This work is partly supported by DGI and FEDER funds, under contract
FIS2006-03438, FIS2008-01143/FIS and PIE-CSIC 200850I238 and the Junta
de Andalucia grant no. FQM225-05. We acknowledge the support of the
European Community-Research Infrastructure Integrating Activity "Study
of Strongly Interacting Matter" (acronym HadronPhysics2, Grant
Agreement n. 227431) under the Seventh Framework Programme of EU.
Work supported in part by DFG (SFB/TR 16, "Subnuclear Structure of
Matter").
\end{theacknowledgments}


\bibliographystyle{aipproc}

\end{document}